\documentclass[preprint]{aastex}
\usepackage{natbib,psfig}

\newcommand{\asca}{{\small \it ASCA}}
\newcommand{\sis}{{\small SIS}}
\newcommand{\gis}{{\small GIS}}
\newcommand{\chip}{{\small CHIP}}
\newcommand{\rosat}{{\small \it ROSAT}}
\newcommand{\hri}{{\small HRI}}
\newcommand{\ginga}{{\it Ginga}}
\newcommand{\sax}{{\small \it BeppoSA$\!$X}}
\newcommand{\cie}{{\small CIE}}
\newcommand{\pie}{{\small PIE}}
\newcommand{\dem}{{\small DEM}}
\newcommand{\ccd}{{\small CCD}}
\newcommand{\faint}{{\small FAINT}}
\newcommand{\bright}{{\small BRIGHT}}
\newcommand{\agn}{{\small AGN}}
\newcommand{\xstar}{{\small XSTAR}}
\newcommand{\mekal}{{\small MEKAL}}
\newcommand{\pexrav}{{\small PEXRAV}}
\newcommand{\xspec}{{\small XSPEC}}
\newcommand{\rrc}{{\small RRC}}
\newcommand{\clr}{{\small CLR}}
\newcommand{\chandra}{{\it Chandra}}

\begin{document}

\title{The Physical Conditions of the X-ray Emission Line Regions \\
       in the Circinus Galaxy}
\author{Masao Sako \altaffilmark{1}, Steven M. Kahn \altaffilmark{1},
        Frits Paerels \altaffilmark{1,2},
        and Duane A. Liedahl \altaffilmark{3}}

\altaffiltext{1}{Columbia Astrophysics Laboratory and Department of Physics,
                 Columbia University, 538 West 120th Street, New York, NY
                 10027; masao@astro.columbia.edu (MS),
                 skahn@astro.columbia.edu (SMK), frits@astro.columbia.edu
                 (FP)}
\altaffiltext{2}{Laboratory for Space Research, Stichting Ruimte Onderzoek
                 Nederland, Sorbonnelaan 2, 3584CA, Utrecht, Netherlands}
\altaffiltext{3}{Department of Physics and Space Technology,
                 Lawrence Livermore National Laboratory,
                 P.O. Box 808, L-41, Livermore,  CA  94550;
                 duane@leo.llnl.gov}

\received{}
\revised{3/23/00}
\accepted{}

\shorttitle{X-ray Emission Lines in the Circinus Galaxy}
\shortauthors{Sako et al.}

\begin{abstract}

  We present a detailed X-ray spectral analysis of the Circinus Galaxy using
  archival data obtained with the \asca\ satellite.  The spectrum shows
  numerous emission lines in the soft X-ray band from highly ionized ions, as
  well as Compton reflection and fluorescent lines from neutral or
  near-neutral matter.  We analyze the spectrum in the context of a
  self-consistent recombination cascade model and find that a nearly flat
  differential emission measure (\dem) distribution in ionization parameter
  fits the data.  For a fixed solid angle distribution of matter surrounding a
  point source, this corresponds to a run of electron density of the form,
  $n(r) \sim r^{-3/2}$, which is suggestive of Bondi accretion onto a central
  compact mass.  Using this density profile and comparing the resulting
  emission spectra with the \asca\ data, the size of the X-ray emission line
  region is estimated to be $\la 1 ~\rm{kpc}$.  We also show that the derived
  density as a function of radius is compatible with the X-ray recombination
  line emission region being the confining medium of the coronal line regions.

\end{abstract}


\section{Introduction}
\label{sec:intro}

  X-ray spectra of Seyfert~2 galaxies show numerous discrete spectral features
  that are produced in a wide variety of physical conditions.  In the few
  cases where the statistical quality of the data has permitted a detailed
  investigation of the spectrum, the soft X-ray band has been found to be
  almost completely dominated by line emission while the hard X-ray emission
  has been found to be consistent with nonthermal radiation possibly produced
  through accretion onto the central compact object.  X-ray observations have
  placed meaningful constraints on the structure of matter immediately
  surrouding the central source, and have played an important role in the
  development of a unified model of \agn\ \citep{antonucci93}.  Observations
  of warm absorbers in Seyfert~1 galaxies, for example, and their response to
  variations in the X-ray continuum provide compelling evidence that a
  significant amount of circumnuclear material is photoionized by the primary
  radiation \citep{otan96}.  There is also observational evidence that a large
  fraction of Seyfert~2 galaxies harbor starburst regions where large amounts
  of mechanically-heated gas are produced, for example, through supernova
  explosions.  Nuclear jets may also heat the surrounding gas through shocks.
  The orgin of the observed soft X-ray emission and the excitation mechanisms
  responsible for producing the observed emission lines in these galaxies,
  however, are not well understood and, in most cases, cannot be discerned
  from the low spectral resolution data available to date.

  In the archetypal Seyfert~2 galaxy NGC~1068, for example, \citet{ueno94}
  find that the soft X-ray spectrum obtained with \asca\ is consistent with
  emission from two-temperature gas in collisional ionization equilibrium
  (\cie), which may be associated with nuclear and spatially extended
  starburst regions seen in the \rosat\ \hri\ data \citep{wilson92}.  On the
  other hand, \citet{netzer97} argue that a combination of warm ($T \sim 1.5
  \times 10^{5} ~\rm{K}$) and hot ($T \sim 3 \times 10^{6} ~\rm{K}$) gas in
  photoionization equilibrium (\pie) is also consistent with the \asca\ data.
  Despite the fact that both models adequately fit the data, their are
  problems in the inferred parameters that cast doubt on these
  interpretations.  The \cie\ fit requires a low-temperature component which
  is essentially metal-free (abundances $\sim 3\%$ of the solar photosphere).
  The \pie\ model requires a steep soft X-ray continuum component (photon
  index $\Gamma = 3.4$) that contains a large fraction of the total X-ray
  luminosity and does not have an obvious physical interpretation.  On the
  other hand, a detailed X-ray spectral analysis of another Seyfert~2 galaxy,
  Mkn~3, by \citet{griffiths98} shows that emission from plasmas both in \cie\
  and \pie\ can fit the composite \ginga, \asca, and \rosat\ data.

  The Circinus Galaxy is the closest ($D \approx 4 ~\rm{Mpc}$;
  \citealt{freeman77}) and one of the brightest Seyfert 2 galaxies in the
  X-ray band.  Extensive studies at various wavelengths have provided firm
  evidence for an obscured Seyfert nucleus at the center of the galaxy; the
  detection of a prominent one-sided [\ion{O}{3}] ionization cone
  \citep{marconi94}, optical and near-IR coronal line emission\footnote{The
  term {\it coronal lines} here refers to high excitation forbidden lines,
  which were first observed in the solar corona.  These lines can, in
  principle, be produced either through photoionization or collisional
  ionization, and are unrelated to emission lines from {\it coronal plasmas},
  a term which refers to optically thin emission from collisionally ionized
  plasmas.} from the nuclear region \citep{oliva94,moorwood96}, bipolar radio
  lobes orthogonal to the plane of the galaxy \citep{elmouttie95}, rapid
  variations of H$_2$O maser emission from the nucleus \citep{greenhill97},
  polarized and broad H$\alpha$ emission from the nucleus \citep{oliva98}, and
  a highly absorbed hard X-ray spectrum with a bright iron K fluorescent line
  near 6.4 keV \citep{matt96b,guainazzi99}.  The system is also known to
  contain regions of starburst activity around the nucleus that account for a
  large fraction of the bolometric luminosity \citep{rowan89}.  Recent studies
  of the infrared spectrum indicate that the relative contributions from
  starburst and \agn\ activity, which can be traced through emission from hot
  dust and ionized gas, respectively, are approximately equal
  \citep{maiolino98,genzel98}.

  In a preliminary analysis of the \asca\ data by \citet{matt96b}, it was
  shown that the 2 -- 10 keV spectrum is dominated by reflection from cold
  matter, possibly originating from the inner wall of the putative molecular
  torus that surrounds the central nucleus.  Several emission lines were also
  detected in the soft X-ray region (0.5 -- 2 keV), most of which were
  inferred to be present in the lower resolution \sax\ spectrum as well
  \citep{guainazzi99}.  The measured energies of these emission lines indicate
  that they are produced through excitation of highly ionized material, and
  cannot be produced in a cold reflection medium.  The soft X-ray region is
  certainly not dominated by cold reflection.

  The purpose of this paper is to perform a detailed analysis of the soft
  X-ray spectrum of the Circinus Galaxy using various spectral models, and to
  derive physical parameters that characterize the structure of the X-ray
  emission line regions.  As we show, the soft X-ray spectrum cannot
  distinguish {\it spectroscopically} between collisionally ionized gas and
  photoionized gas \citep[cf.,][]{netzer98}.  There is a strong piece of
  evidence, however, against an interpretation in terms of collisionally
  ionized gas, and the inferred parameters somewhat favor photoionization
  equilibrium.  We discuss the implications of our results and investigate the
  properties of the active nucleus and its circumnuclear environment.

\section{Observation and Data Reduction}
\label{sec:data_redu}

  The Circinus Galaxy was observed with \asca\ during August 14 -- 15, 1995.
  For the purposes of our analysis, we only use data obtained by the
  Solid-state Imaging Spectrometers (\sis), which have the highest spectral
  resolving power in the soft X-ray band.  The \sis\ data were taken in
  \faint\ and \bright\ 2-\ccd\ modes, while most of the photons from the
  Circinus Galaxy were recorded on \sis0-\chip1 and \sis1-\chip3.  A quick
  analysis of the \sis0 and \sis1 data indicated a $\sim 1\%$ offset in the
  energy scale in the iron-K region, while no obvious offset was observed in
  the soft X-ray band (although this is most likely due to the poor
  statistical quality of data in that region).  We applied a $\sim 1\%$ gain
  correction to the \sis1-\chip3 response matrix.  Since the spectral
  resolving power of the \gis\ is lower by a factor of $\sim 3$ in the soft
  X-ray region, the discrete features observed in the \sis\ are smeared out in
  the \gis.  We, therefore, only use data collected on the \sis\ for our
  spectral analysis.

  The data were screened though the standard criteria using {\small FTOOLS}
  v4.0.  After conversion of \faint\ mode data to \bright\ mode data, we
  combined them with true \bright\ mode data to obtain a total exposure time
  of 33 ksec amounting to $\sim 4000$ and $\sim 3000$ counts on the \sis0 and
  \sis1 detectors, respectively.  Although this method does not fully exploit
  the resolving power capabilities of the \sis, we can significantly improve
  the statistical quality of the data by compromising for a 20 -- 30\%
  degredation of the resolving power.

\section{Spectral Analysis}
\label{sec:spec_anal}

  In this section, we discuss the spectral models that we use for our analysis
  of the \sis0 data.  All of the models we consider here consist of (1) a
  reflection component, (2) a soft continuum component, and (3) a soft X-ray
  emission line component.  Possible physical interpretations of each of the
  spectral components are as follows.  The reflection component represents a
  fraction of the intrinsic \agn\ continuum radiation that is
  Compton-scattered in a cold medium into our line of sight.  The spectrum
  resembles that of a highly absorbed continuum with a number of K-shell
  fluorescent lines superimposed \citep{lightman88,george91,matt96a}, which we
  simply represent by a set of gaussian lines and a Compton reflected powerlaw
  continuum (i.e., \pexrav\ in \xspec).  This component has a number of free
  parameters; line energies and fluxes of K-shell Ne, Mg, Si, S, Ar, Ca, and
  Fe, the intrinsic powerlaw photon index, and the total reflected flux.  We
  assume that the K-shell fluorescent lines are infinitely narrow, which is
  justified by the observed Fe K$\alpha$ line profile, as we discuss later.
  The soft continuum component may be produced through electron scattering of
  the primary radiation in the surrounding gas, in which case the photon index
  should be equal to that of the primary radiation, or through inverse Compton
  scattering of infrared photons with relativistic electrons.  For the latter
  case, the photon index should match the photon index in the radio band.  The
  soft X-ray emission line component is modeled as either a collisionally
  ionized or a photoionized plasma in equilibrium.  We assume that the entire
  spectrum is absorbed by a single column density (Galactic + local
  absorption).  The reflection component possibly suffers higher absorption,
  since the reflection region (the inner wall of the molecular torus) is
  assumed to be located closer to the nucleus.  The reflected spectrum,
  however, is not sensitive to $N_H \la 10^{22} ~\rm{cm}^{-2}$ since most of
  the flux is contained in the region above $\sim 2 ~\rm{keV}$ anyway.
  Throughout this paper, we adopt a distance of 4~Mpc to the Circinus Galaxy.

\subsection{Collisional Ionization Equilibrium Model}
\label{subsec:cie}

\begin{figure}
  \centerline{\psfig{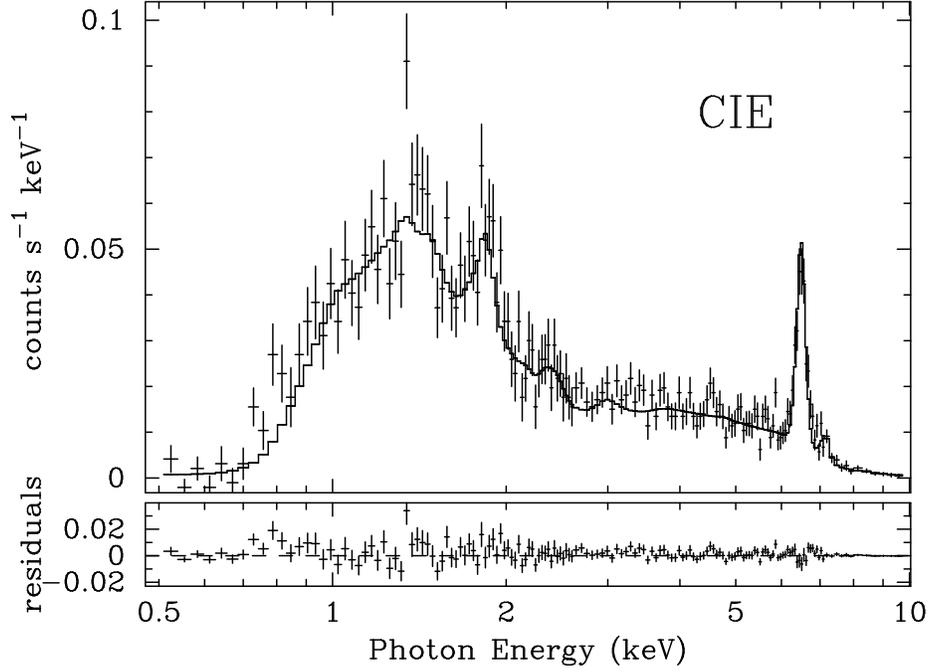}}
  \caption{The \asca\ \sis\ data and the best-fit \cie\ model.  The fit is
           statistically acceptable with $\chi^2_r = 1.21$ for 305 degrees of
           freedom.  Only \sis0 data is shown for clarity.}  \label{f1}
\end{figure}

  We first adopt a model in which the soft X-ray component is represented by a
  single temperature thermal plasma in collisional ionization equilibrium
  (i.e., \mekal\ in \xspec), with the temperature, overall abundance, and the
  emission measure free parameters in the fit.  For the reflection component,
  we first allowed the fluorescent line energies to be free, and found that
  the model is unable to locate the centroid energies for all but the Fe
  K$\alpha$ and K$\beta$ lines.  We, therefore, froze the energies of all of
  the fluorescent lines, except for those of iron, at values shown in
  Table~\ref{tbl1}, which are those derived from the photoionization
  equilibrium fit discussed in \S~\ref{subsec:pie}.  These are more or less
  consistent with the ionization state inferred from the iron K$\alpha$ line
  energy (i.e., partially ionized medium), which is also consistent with the
  iron K$\alpha$ line energy measured with \sax\ \citep{guainazzi99}.  The
  measured fluorescent line intensities are also listed in Table~\ref{tbl1}.
  Contrary to \citet{matt96b}, we find a statistically acceptable fit with
  $\chi^2_r = 1.21$ for 305 degrees of freedom.  The best-fit model in
  comparison with the data is shown in Figure~\ref{f1} and the derived
  parameters are listed in Table~\ref{tbl2}.

\placetable{tbl1}
\placefigure{f1}
\placetable{tbl2}

\begin{figure}
  \centerline{\psfig{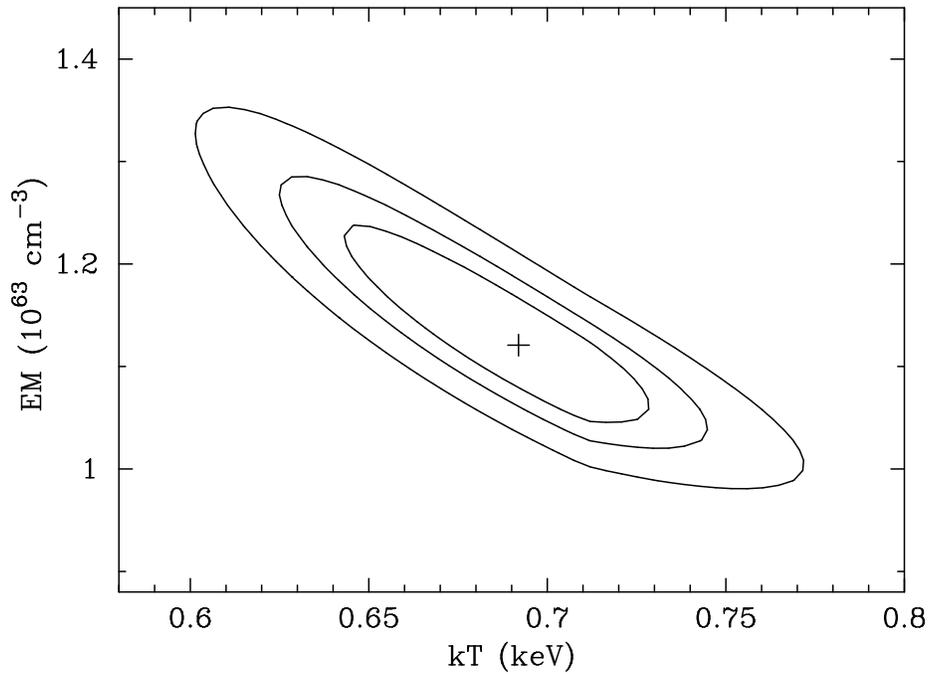}}
  \caption{Two parameter confidence ranges ($kT$ vs. total $EM$) for the
     highly ionized component in the \cie\ fit.  Solar abundances are
     assumed.}  \label{f2}
\end{figure}

  The allowed ranges in temperature and emission measure ($EM = \int
  n_e^2~dV$) are shown in Figure~\ref{f2} where we plot 68, 90, and 99\%
  confidence ranges for two interesting parameters.  We obtain 90\% confidence
  ranges of $kT = 0.69^{+0.05}_{-0.07}$ keV and $EM = (1.12^{+0.17}_{-0.10})
  \times 10^{63} ~\rm{cm}^{-3}$.  Including an additional thermal component
  does not improve the fit significantly ($\Delta \chi^2 \sim 1$).  In fact,
  if we adopt a powerlaw differential emission measure (\dem) model where the
  emission measure as a function of $kT$ is proportional to $(kT)^{\alpha}$,
  we obtain a steep slope of $\alpha \sim 5$ and a sharp high-temperature
  cutoff at $kT \sim 0.7$ keV, which nearly corresponds to a single
  temperature model at $kT \sim 0.7$ keV.  Similar results are obtained for
  different \dem\ models in coronal equilibrium.  In any case, the total
  emission measure in the thermal component is $\sim 1 \times 10^{63}
  ~\rm{cm}^{-3}$ and the corresponding unabsorbed luminosity is $L_x = 3
  \times 10^{40} ~\rm{erg ~s}^{-1}$.

\placefigure{f2}

\subsection{Photoionization Equilibrium Model}
\label{subsec:pie}

  We next test a model in which the soft X-ray emission lines are produced
  through cascades following radiative recombination in photoionization
  equilibrium.  This spectral model was originally constructed for the
  analysis of the \asca\ spectrum of Cygnus X-3 \citep{liedahl96} and was also
  succesfully applied to the \asca\ data of Vela X-1 \citep{sako99}.  The
  model includes line emission from hydrogen- and helium-like carbon,
  nitrogen, oxygen, neon, magnesium, silicon, sulfur, argon, and calcium, as
  well as the highest ten charge states of iron (\ion{Fe}{17} --
  \ion{Fe}{26}), and the associated radiative recombination continua (\rrc).
  For a detailed description of the atomic model, see \citet{sako99}.

  We first start by fitting the data with the recombination cascade model,
  using only the K-shell ions of oxygen, neon, magnesium, silicon, sulfur,
  argon, calcium, and iron, to construct an empirical \dem.  Due to the
  limited statistical quality of the data, we cannot directly constrain the
  temperature of each ion emission region from the shape of the \rrc, so we
  fix the temperature of each ion at their respective temperature of
  formation.  We do this by first calculating the ionization balance and
  temperature as functions of the ionization parameter ($\xi = L_x/(n_pr^2)$,
  where $L_x$ is the X-ray luminosity between 1 and 100 Rydbergs, $n_p$ is the
  proton number density, and $r$ is the distance from the ionizing source;
  \citealt{tarter69}) using the photoionization code, \xstar\
  \citep{kallman95}.  We assume that the medium is optically thin with respect
  to continuum absorption, which is an excellent approximation, as we show
  later.  We adopt an exponential cutoff powerlaw ionizing continuum with a
  photon index of $\Gamma = 1.7$ with an e-folding energy of $E_f = 100
  ~\rm{keV}$ and a cutoff energy of $E_c = 60 ~\rm{keV}$ \citep{guainazzi99}.
  Using these results, we then calculate the temperature at which each of the
  ion line emissivities is a maximum and fix each ionic temperature at those
  values, which we refer to as the temperatures of formation.  This is a good
  first order approximation since a large fraction of the total ion line
  fluxes are produced near those temperatures.  Approximately 50\% of the
  recombination line emissivity of the He-like ions originates from a range in
  $\Delta \log \xi$ of $\sim 0.5$.  For the H-like ions, the width is broader
  ($\Delta \log \xi \sim 1$) since the emission lines are formed through
  recombination onto a bare nucleus, which exists in a wider range in
  ionization parameter.

\begin{figure}
  \centerline{\psfig{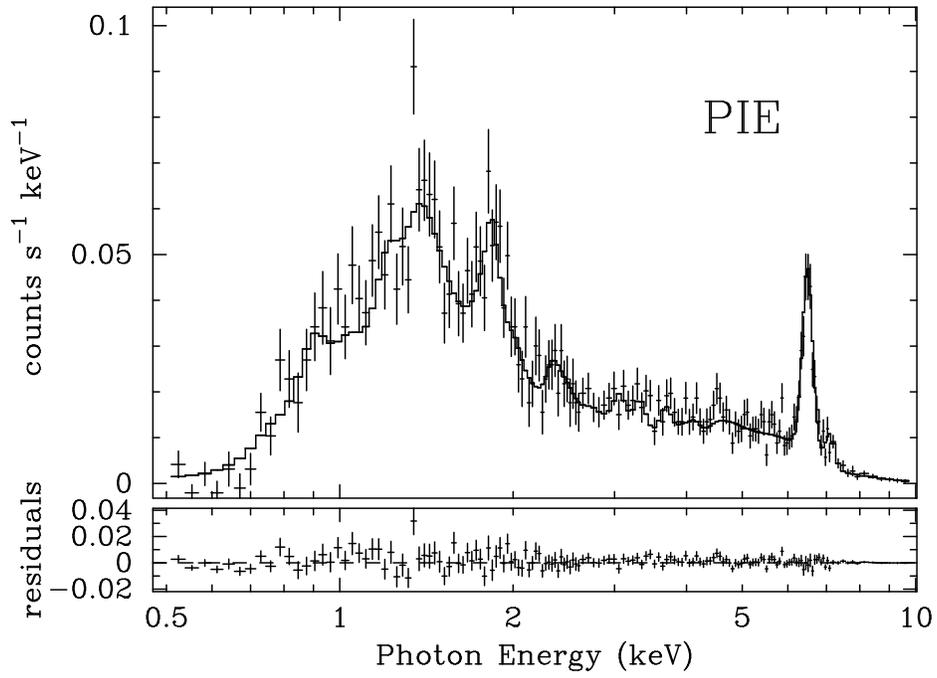}}
  \caption{The best-fit recombination cascade model compared with the \sis\
     data ($\chi_r^2 = 1.10$ with 290 degrees of freedom).  Again, only \sis0
     data is shown for clarity.} \label{f3}
\end{figure}

\begin{figure}
  \centerline{\psfig{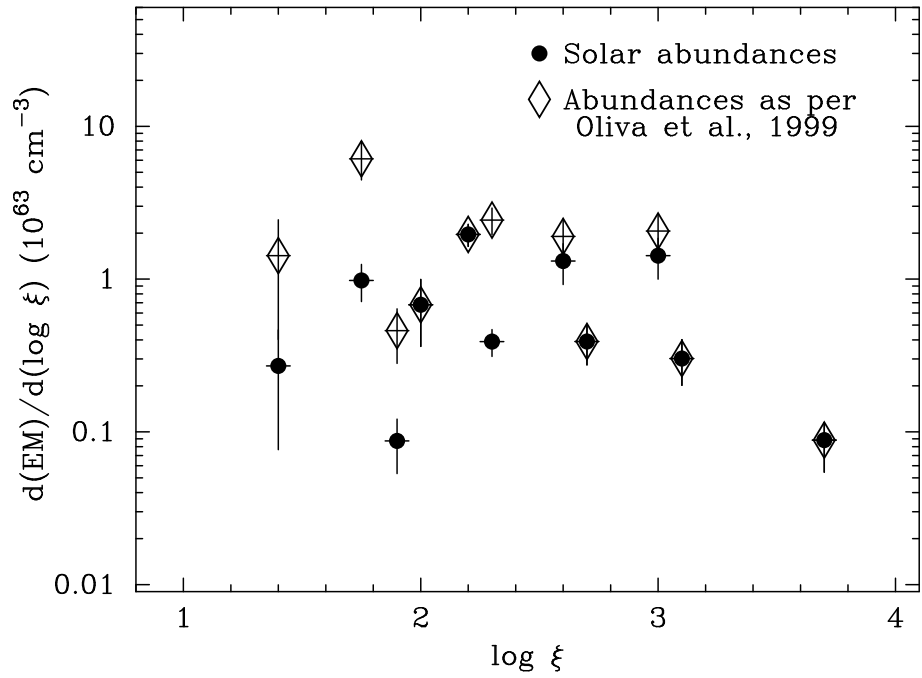}}
  \caption{An empirical \dem\ distribution derived from K-shell oxygen, neon,
     magnesium, silicon, argon, and iron for two sets of assumed chemical
     abundances; solar photospheric abunandes of \citet{anders89} and
     abundances derived by \citet{oliva99} for a knot located near the nucleus.}
     \label{f4}
\end{figure}

  We fit the \sis0 and \sis1 data simultaneously and obtain a statistically
  acceptable fit with $\chi_r^2 = 1.10$ with 290 degrees of freedom.  The data
  in comparison to the model are plotted in Figure~\ref{f3}.  In addition to
  the continuum parameters, this model contains normalizations of each ion
  line flux, which are proportional to their ion line emission measures.
  Figure~\ref{f4} shows the derived empirical \dem\ distribution assuming an
  emission width of $\Delta \log \xi \sim 0.5$ for helium-like ions and
  $\Delta \log \xi \sim 1$ for the hydrogenic species for two sets of assumed
  chemical abundances; (1) solar photospheric abundances \citep{anders89} and
  (2) abundances derived by \citet{oliva99} for a knot located approximately
  $15\arcsec$ from the nucleus in the direction of the [\ion{O}{3}] cone.  We
  adopt solar abundances for elements that are not listed in \citet{oliva99}.
  Given the assumptions that we have made in the construction of the empirical
  \dem\ distribution, we estimate the points drawn on Figure~\ref{f4} to be
  uncertain by up to a factor of 2 -- 3.

\placefigure{f3}
\placefigure{f4}

  With the exception of the \dem\ points derived from \ion{O}{7} and
  \ion{O}{8} at $\log \xi = 1.4$ and $\log \xi = 1.9$, respectively, the shape
  of the empirical \dem\ is nearly linear with evidence of a slight decrease
  in \dem\ with increasing $\log \xi$.  The oxygen ions have bright features
  below $E \sim 1 ~\rm{keV}$ and the derived emission measures are extremely
  sensitive to the foreground column density.  An increase in the column
  density by $50\%$ can be offset by an increase in the emission measure of
  \ion{O}{7} by a factor of $\sim 10$ and \ion{O}{8} by a factor of $\sim 6$.
  Keeping this in mind, the derived \dem\ distribution, to first order, is
  nearly linear.

\begin{figure}
\centerline{\psfig{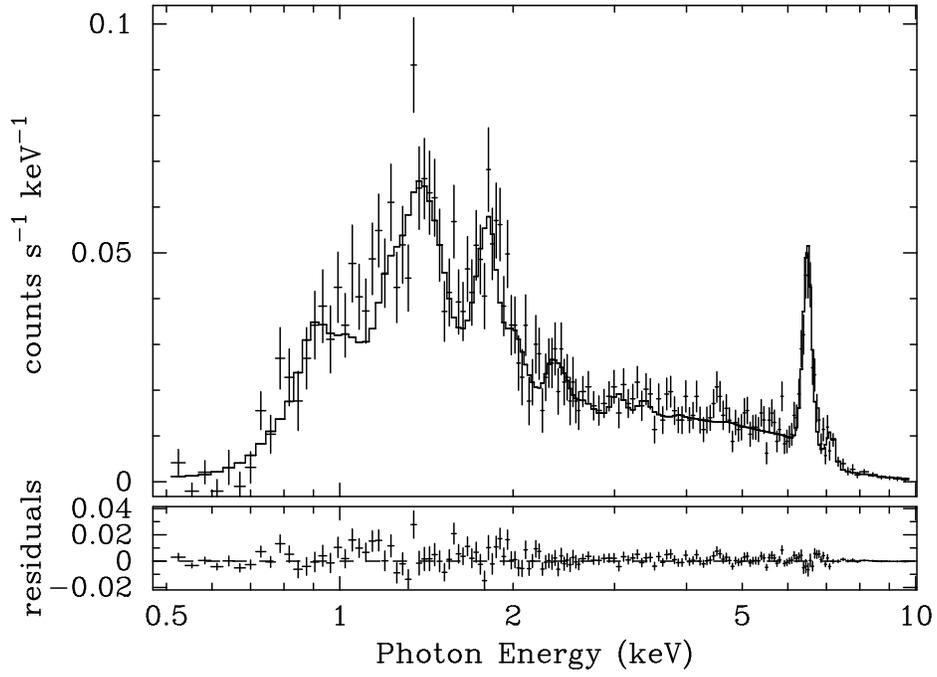}}
  \caption{The best-fit powerlaw \dem\ \pie\ model in comparison to the data.
      Again, the fit is acceptable with $\chi_r^2 = 1.12$ for 299 degrees of
      freedom.}  \label{f5}
\end{figure}

\begin{figure}
  \centerline{\psfig{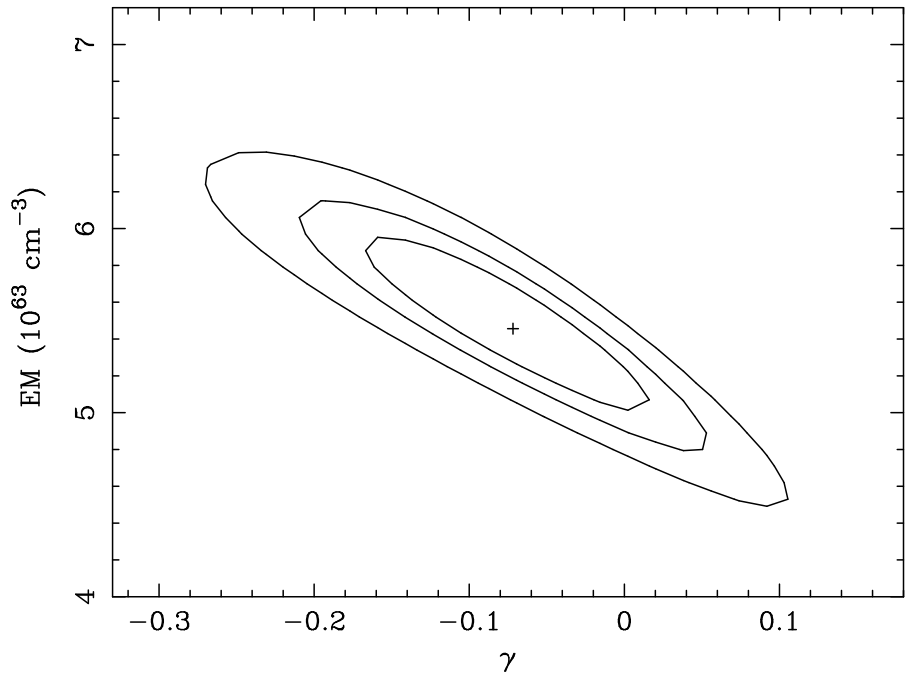}}
  \caption{Two parameter confidence ranges for the slope of the \dem\ curve,
     $\gamma$, versus the total emission measure, $EM$, for the \pie\ \dem\
     fit.}  \label{f6}
\end{figure}

\begin{figure}
  \centerline{\psfig{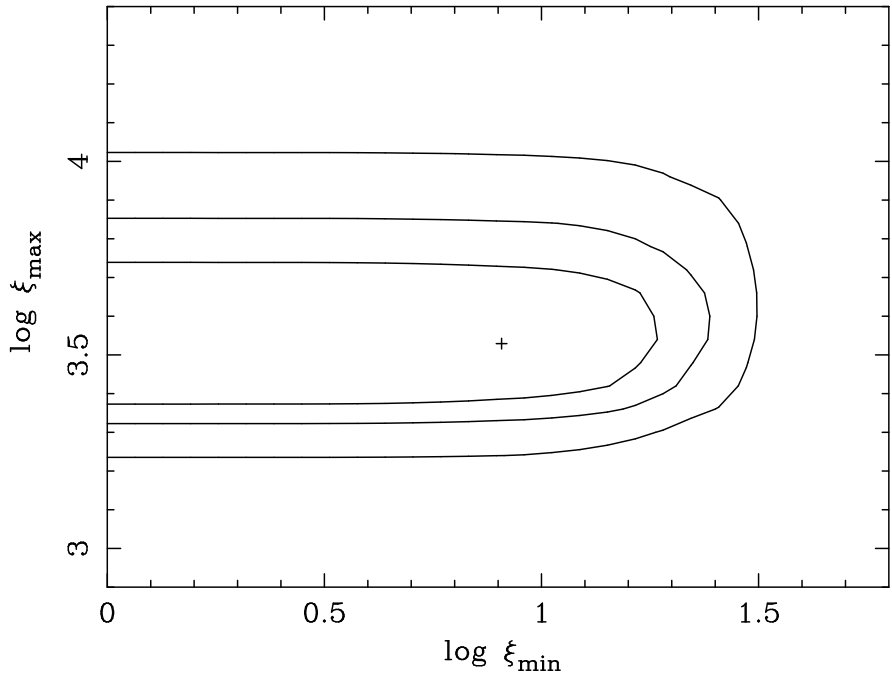}}
  \caption{Two parameter confidence ranges for the lower and upper cutoff
     ionization parameters ($\log \xi_{\rm{min}}$ vs. $\log \xi_{\rm{max}}$)
     for the \pie\ \dem\ fit.}  \label{f7}
\end{figure}

\begin{figure}
  \centerline{\psfig{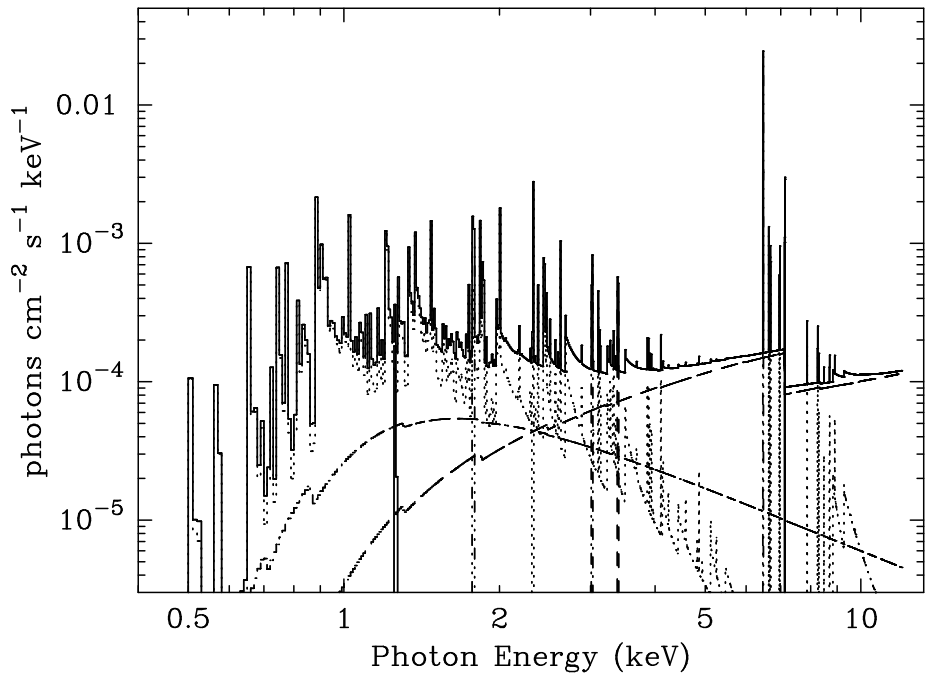}}
  \caption{The best-fit \pie\ \dem\ model spectrum for the parameters shown in
     Table~\ref{tbl4}.}  \label{f8}
\end{figure}

  We then adopt a reverse fitting procedure where we parametrize the model
  spectrum of the highly ionized component according to powerlaw \dem s of the
  form, $d(EM)/d \log \xi \propto \xi^{\gamma}$, with minimum and maximum
  cutoff ionization parameters, $\xi_{\rm{min}}$ and $\xi_{\rm{max}}$,
  respectively.  This procedure bypasses the ambiguities in the assumed
  temperatures of formation and correctly takes account of the temperature
  gradient in each ionization zone.  In addition to the K-shell ions that we
  had initially used in the empirical fit, this time we also include line and
  \rrc\ emission from the iron L species, which are expected to contribute
  $\sim 10\%$ to the total recombination emission flux.  There are four free
  parameters in this component, namely, the slope $\gamma$ of the \dem\ as a
  function of ionization parameter, the normalization, which is proportional
  to the total emission measure $EM$, low-$\xi$ cutoff $\xi_{\rm{min}}$, and
  high-$\xi$ cutoff $\xi_{\rm{max}}$.  Unlike the \cie\ case, we are able to
  centroid the fluorescent line energies of all the elements that we list in
  Table~\ref{tbl2}, and we find a statistically acceptable fit with $\chi_r^2
  = 1.12$ for 299 degrees of freedom.  The data in comparison to the model is
  shown in Figure~\ref{f5} and the best-fit parameters are listed in
  Table~\ref{tbl3}.  We also derive confidence contours for two pairs of
  interesting parameters; ($\xi_{\rm{min}}$, $\xi_{\rm{max}}$) and ($\gamma$,
  $EM$), as shown in Figures~\ref{f6} and~\ref{f7}.  In summary, we derive
  90\% confidence ranges of $\gamma = -0.07^{+0.12}_{-0.14}$ and $EM = (5.5
  \pm 0.7) \times 10^{63} ~\rm{cm}^{-3}$ for two interesting parameters, and
  $\log \xi_{\rm{min}} = 0.91^{+0.48}_{-\infty}$ and $\log \xi_{\rm{max}} =
  3.53^{+0.32}_{-0.20}$, again, for two interesting parameters.  The best-fit
  model spectrum with the various components is shown in Figure~\ref{f8}.

\placefigure{f5}
\placefigure{f6}
\placefigure{f7}
\placetable{tbl3}
\placetable{tbl4}
\placefigure{f8}

  The lower limit on $\xi_{\rm{min}}$ cannot be well-constrained simply
  because the X-ray spectrum of the Circinus Galaxy is absorbed through a
  moderately high column density ($N_H \sim 8 \times 10^{21} ~\rm{cm}^{-2}$),
  and also because the \asca\ \sis\ detectors are not sensitive to X-rays
  below $\sim 0.5$~keV, where discrete recombination emission from regions
  with $\log \xi \sim 1$ dominates, for example, from carbon and nitrogen.
  The upper limit on $\xi_{\rm{max}}$ is constrained mainly by the Ly$\alpha$
  flux of \ion{Fe}{26}.

\section{Discussion and Implications}

  We have shown that the soft X-ray emission line spectrum of the Circinus
  Galaxy is compatible with both coronal and photoionization equilibrium, and
  the excitation mechanisms cannot be spectroscopically distinguished without
  any ambiguity.  In this section, we extract physical parameters implied by
  each model and discuss their validity in relation to observations at other
  wavelength bands.

  According to recent \rosat\ and \asca\ observations, nearby star-forming
  galaxies show X-ray spectra that can be well described by a combination of a
  soft optically thin thermal plasma with a temperature of $kT \sim 0.6
  ~\rm{keV}$ and a hard continuum component with either a photon index of
  $\Gamma \sim 1.7$ or a bremsstrahlung temperature of $kT \sim 5 ~\rm{keV}$
  (e.g., \citealt{ptak99}, and references therein).  The collisional
  ionization interpretation for the Circinus Galaxy gives a best-fit
  temperature of $\sim 0.69 ~\rm{keV}$ and a luminosity of $\rm{few} \times
  10^{40} ~\rm{erg ~s}^{-1}$ in this component, which are typical values for
  starburst galaxies.  Note that the temperature of the plasma is
  well-constrained and that the data do not allow the presence of gas with
  temperature lower than $kT \sim 0.60 ~\rm{keV}$ at the 99\% confidence
  level.  It is interesting to compare this with what \citet{oliva94} have
  found using visible and near infrared observations of the coronal emission
  lines in the Circinus Galaxy.  In order to reproduce the observed spectrum
  in the context of a powerlaw \dem\ plasma in \cie, they require a sharp
  cutoff in the \dem\ at a temperature of $kT_{\rm{max}} \sim 0.1 ~\rm{keV}$.
  This would imply that there is a gap in the emission measure distribution in
  the range $0.1 \la kT \la 0.6 ~\rm{keV}$, which has no obvious physical
  interpretation.

  In \pie, the assumption of plasma heating by a point source of X-radiation
  places an additional constraint on the plasma structure (e.g., the heating
  rate as a function of $\xi$), which can be then used to derive the
  geometrical distribution of the gas.  In the \pie\ interpretation of the
  spectrum, have shown that a nearly flat \dem\ distribution is consistent
  with the data.  One simple way of producing a flat \dem\ curve is to adopt a
  radial density profile of the form, $n_e(r) \propto r^{-3/2}$.  This is
  perhaps the simplest configuration but, by no means, unique.  More
  generally, a powerlaw density profile produces a powerlaw \dem\ curve.  It
  is interesting to note, however, that a conical homogeneous medium with a
  powerlaw density profile as a function of radius always produces a powerlaw
  \dem\ distribution; i.e., a simple calculation shows that for a density
  profile of the form,
\begin{equation}
  \label{density}
  n_e(r) = n_{e0}~r_{\rm{pc}}^{-\beta},
\end{equation}
  where we define $n_{e0}$ to be the density in cm$^{-3}$ at a distance of
  1~pc from the central source, the \dem\ distribution can be written as,
  $d(EM) = \Delta \Omega ~n_e^2 ~r^2 ~dr$, or
\begin{equation}
  \label{dem}
  \frac{d(EM)}{d(\log \xi)} = \ln 10 ~\frac{\Delta \Omega}{|\beta - 2|}
    ~[n_{e0} (\mu L_x)^{3-2\beta}]^{1/(2-\beta)}
    ~\xi^{(3-2\beta)/(\beta-2)},
\end{equation}
  where $\mu$ is the ratio of the electron to proton number densities ($\mu =
  1.2$, for an X-ray emitting plasma) and $\Delta \Omega$ is the opening solid
  angle of the photoionized region.  From this equation, we can identify the
  slope of the \dem\ curve $\gamma$ to be $(3-2\beta)/(\beta-2)$.  Our
  inferred range in $\gamma$ corresponds to $\beta = 1.48^{+0.03}_{-0.04}$.
  One can also see that $\beta = 3/2$ corresponds to a flat \dem\ curve.

  The fact that $\beta$ is relatively well-constrained is due to the fact that
  the slope of the \dem\ curve is extremely sensitive to the value of $\beta$.
  As mentioned earlier, a $\beta = 3/2$ density profile produces a flat \dem\
  curve, which produces emission lines and \rrc\ from a wide range of
  ionization stages.  A $\beta = 2$ profile, on the other hand, corresponds to
  a single-$\xi$ zone, since both the ionizing flux and the density decay like
  $\propto r^{-2}$.  In this case, only a small number of charge states can
  exist, and the resulting spectrum is more sparse.

  Our derived values for the parameters shown in Table~\ref{tbl4} depend on
  the assumed metal abundances.  If we assume that the abundances are
  $A_{Z_{\sun}}$ relative to solar, the \dem\ curve will be uniformly shifted
  by a factor of $A_{Z_{\sun}}^{-1}$ and the inferred total emission measure
  will be $EM = 5.5 \times 10^{63} ~A_{Z_{\sun}}^{-1} ~\rm{cm}^{-3}$.  The
  parameters, $\beta$, $\xi_{\rm{min}}$, and $\xi_{\rm{max}}$ are unaffected
  by a uniform shift in abundances, since they characterize only the shape of
  the \dem\ curve and not its normalization.  The exact dependences of these
  parameters on the abundance of an individual element, however, is more
  complex.  For example, an underabundance of oxygen and neon, as found by
  \citet{oliva99}, may result in a steeper \dem\ slope ($\beta \ga 1.5$) or a
  decrease in $\xi_{\rm{min}}$ in order to increase the emission measure at
  low-$\xi$ and compensate for the observed flux at low energies.

  Although the density parametrization of equation~\ref{density} is
  simplistic, chosen for the photoionization interpretation of the spectrum,
  there is empirical evidence that narrow-line regions in many Seyfert
  galaxies, in fact, do exhibit this type of behavior.  For $\beta = 2$, the
  density scales like $r^{-2}$, and, hence, the ionization parameter is
  constant with radial distance.  This has been observed in some systems that
  show correlations between widths of optical emission lines and their
  critical densities for collisional de-excitation.  Spectral transitions with
  higher critical densities generally exhibit broader line profiles.  There
  are also cases where the observed widths are higher for transitions in ions
  with higher ionization potential, which corresponds to a scenario with
  $\beta < 2$ (e.g., \citealt{pelat81,filippenko84,derobertis84}).  According
  to \citet{derobertis86}, values of $\beta$ lie between 0 and 2, while more
  Seyfert 2 galaxies have $\beta$ closer to 2.  A $\beta > 2$ behavior (i.e.,
  $\xi$ increasing with distance) is not observed in any Seyfert galaxy.

  Using the empirical density profile, we can estimate several physical
  parameters from the results of the \dem\ fit.  From the best-fit values for
  $\beta$ and the total $EM$, which is proportional to the prefactor in
  equation~\ref{dem}, $\Delta \Omega (n_{e0} L_x^{3-2\beta})^{1/(2-\beta)}$,
  we estimate the density profile of the X-ray emission region to be,
\begin{equation}
  \label{density2}
  n_e(r) = 4.0 \times 10^{2}
     ~\Delta \Omega^{-0.52} ~L_{x(42)}^{-0.04}
     ~r_{\rm{pc}}^{-1.48} ~(\rm{cm}^{-3}),
\end{equation}
  where $L_{x(42)}$ is the X-ray luminosity in multiples of $10^{42}
  ~\rm{erg~s}^{-1}$.  We also saw that the upper and lower limits on the
  ionization parameter were $\log \xi_{\rm{max}} \sim 3.5$ and $\log
  \xi_{\rm{min}} \la 0.9$, respectively.  From the upper limit, we can
  estimate the distance from the central nucleus to the inner-most X-ray
  emission line region that is directly visible from the observer.  We
  estimate the inner distance to be $r_{\rm{in}} \sim 0.011 ~\Delta \Omega
  ~L_{x(42)}^{2} ~\rm{pc}$.  Similarly, the lower limit on the outer distance
  is $r_{\rm{out}} \sim 1200 ~\Delta \Omega ~L_{x(42)}^{2} ~\rm{pc}$.  We can
  see from the [\ion{O}{3}] image that the opening half-angle of the one-sided
  ionization cone is $\sim 30\degr$, which corresponds to a solid angle of
  $\Delta \Omega \sim 0.8$.  Assuming that the X-ray emission lines are
  produced in the inner region close to the base of this ionization cone, we
  estimate its size to be $\sim 900 ~\rm{pc}$ or approximately $50 \arcsec$ in
  the sky.  This is qualitatively consistent with the \rosat\ \hri\ image that
  shows a slight spatial extent ($\la 30 \arcsec$) approximately in the
  direction of the [\ion{O}{3}] cone.

  As a cross-check on the derived parameters, we calculate the electron
  scattering optical depth through the emission line medium and compare it
  with the observed scattered continuum flux.  As listed in Table~\ref{tbl3},
  the unabsorbed luminosity of the scattered continuum is $L_x^{\rm{scat}}
  \sim 8.4 \times 10^{39} ~\rm{erg~s}^{-1}$, which corresponds to a Thomson
  optical depth of $\tau_{\rm{es}} = 0.084 ~(\Delta \Omega ~L_{x(42)})^{-1}$.
  Integrating the density profile (equation~\ref{density2}) from $r_{\rm{in}}$
  out to $r_{\rm{out}}$ yields an optical depth of, $\tau_{\rm{es}} = 0.014
  ~(\Delta \Omega L_{x(42)})^{-1}$.  We note that regions beyond
  $r_{\rm{out}}$ do not contribute significantly to the optical depth.  This
  implies the presence of additional scattering regions in the circumsource
  medium, which must be either more highly ionized or less ionized than the
  X-ray emission line region.  The additional column density required is $N_H
  ~(\Delta \Omega/4\pi) \sim 1 \times 10^{23} ~\rm{cm}^{-2}$, which may be
  produced in the region that absorbs the transmission component observed in
  the hard X-rays.  The measured column density of this component is $N_H \sim
  4 \times 10^{24} ~\rm{cm}^{-2}$ \citep{guainazzi99}, and thus a covering
  fraction of $(\Delta \Omega/4 \pi) \sim 0.03$ can produce most of the
  observed soft continuum flux.  An optical depth of $\tau_{\rm{es}} \sim
  0.01$ through the photoionized gas amounts to a total column density of $N_H
  \sim ~\rm{few} \times 10^{21} ~\rm{cm}^{-2}$, which justifies our assumption
  that the primary radiation is optically thin to continuum absorption.

  It is interesting to note that a density profile of the form $n(r) \sim
  r^{-3/2}$ is expected for Bondi accretion in a spherically symmetric
  gravitational potential of a point source.  For a central black hole mass of
  $M_{6}$ (in multiples of $10^{6} ~M_\sun$), the gas is nearly in free-fall
  within an accretion radius of $r_{\rm{acc}} \sim 100 ~M_6
  ~T_{\infty(4)}^{-1} ~\rm{pc}$, where $T_{\infty(4)}$ is the ambient gas
  temperature (in multiples of $10^4 ~\rm{K}$) far away from the central
  compact object.  Using the density profile of equation~\ref{density2} and a
  free-fall velocity profile (i.e., $v \propto r^{-1/2}$), the mass inflow
  rate can be expressed as,
\begin{equation}
  \label{mdot_dem}
  \dot{M}_{\rm{inflow}} = 8 \times 10^{-4} ~\Delta \Omega^{0.48}
            ~L_{x(42)}^{-0.04} ~M_{6}^{1/2} ~(M_\sun ~\rm{yr}^{-1}).
\end{equation}
  The mass accretion rate implied by the observed X-ray luminosity is,
\begin{equation}
  \label{mdot_lumin}
  \dot{M}_{\rm{accretion}} = 2 \times 10^{-4} ~L_{x(42)} ~\eta_{0.1}^{-1}
            ~(M_\sun ~\rm{yr}^{-1}),
\end{equation}
  where $\eta_{0.1}$ is the energy conversion efficiency in multiples of
  $0.1$.  Since the accretion radius is on the order of or larger than the
  size of the X-ray emission line region, the outer region may not necessarily
  be inflowing at the free-fall velocity, especially since photoelectric
  heating by the \agn\ continuum radiation increases the local sound speed of
  the surrounding gas.  Nevertheless, these mass flow rates are roughly
  consistent with each other and we suggest that the inner part of the X-ray
  emission region is fueling the central engine.  Projected velocity shifts of
  up to $\sim 230 ~\rm{km~s}^{-1}$ have also been detected near the base of
  the [\ion{O}{3}] and H$\alpha$ ionization cones, which may be interpreted as
  either an outflow or inflow relative to the nucleus
  \citep{veillleux97,elmouttie98}.

  We suggest that the region that produces the observed X-ray recombination
  lines is the confining medium of the so-called coronal line regions
  (\clr;~\citealt{oke68}).  Using the surface brightness of [\ion{Fe}{11}],
  \citep{oliva94} finds the \clr\ to have a size $\sim 0.5\arcsec$ (diameter
  of $\sim 10 ~\rm{pc}$) and a gas density of $n_e \sim 250 ~f^{-1/2}
  ~\rm{cm}^{-3}$, where $f$ is the volume filling factor.  The temperature is
  assumed to be $kT \sim 3 \times 10^{4} ~\rm{K}$, a typical value for a
  photoionized \clr.  Using the density profile inferred from our fit, we
  estimate the gas density at a distance of $\sim 10 ~\rm{pc}$ to be $n_e \sim
  10 ~\rm{cm}^{-3}$.  According to our model, the ionization parameter at this
  distance is $\log \xi \sim 2.5$ with a temperature of $kT \sim 2 \times
  10^{6} ~\rm{K}$, and thus the pressure is $n_e T \sim 2 \times 10^{7}$.  In
  order for the two regions to be in pressure equilibrium, the filling factor
  of the \clr\ must be $f \sim 0.1$.  This implies the density of the
  [\ion{Fe}{11}] region to be $n_e \sim 800 ~\rm{cm}^{-3}$, which is still
  much lower than the critical density of the upper level and, therefore, the
  line will not be diminished.

  The model that we have suggested can be further tested with high spectral
  and spatial resolution grating observations with the \chandra\ observatory.
  Our model predicts that $\sim 70\%$ of the total flux will be spatially
  unresolved, containing the reflection component, as well as a large fraction
  of the scattered soft continuum component.  The remaining $\sim 30\%$, which
  is dominated by recombination line emission, will be extended.
  Specifically, recombination lines from low-$Z$ ions (e.g., O, Ne, and Mg)
  will be extended by $\sim 10\arcsec$ and those from higher-$Z$ ions (e.g.,
  Si, S, Ar, Ca, and Fe) will have a spatial extent of $\la 5\arcsec$.  In
  addition to the spatial distibution of the X-ray emission line region,
  measurements of velocity shifts will allow us to constrain the kinematics
  and determine whether the confining medium is flowing towards or away from
  the central nucleus.

\acknowledgements

  This work has benefited from useful discussions with Andy Rasmussen
  concerning the \asca\ \sis\ calibrations.  We have made use of data obtained
  through the High Energy Astrophysics Science Archive Research Center Online
  Service, provided by the {\small NASA}/Goddard Space Flight Center.  We
  thank the \asca\ team and the {\small HEASARC} Online Service for their
  support.  This work was supported under a {\small NASA} Long Term Space
  Astrophysics Program grant ({\small NAG}5-3541).  D. A. L. was supported in
  part by a {\small NASA} Long Term Space Astrophysics Program grant ({\small
  S}-92654-{\small F}).  Work at {\small LLNL} was performed under the
  auspices of the U. S. Department of Energy, Contract No. {\small
  W}-7405-Eng-48.

\clearpage

\begin{deluxetable}{lcccccc}
\tablecolumns{3}
\tablewidth{0pt}
\tablecaption{Summary of Fluorescent Line Intensities \label{tbl1}}
\tablehead{
  \colhead{} & & \multicolumn{2}{c}{\cie\ Interpretation\tablenotemark{a}} &
  \multicolumn{2}{c}{\pie\ Interpretation\tablenotemark{a}} \\
  \cline{3-4} \cline{5-6} \\
  \colhead{Element} & \colhead{Line Energy\tablenotemark{b} (keV)} &
  \colhead{Line Intensity\tablenotemark{c}} &
  \colhead{$\Delta \chi^2$} &
  \colhead{Line Intensity\tablenotemark{c}} &
  \colhead{$\Delta \chi^2$} &
  \colhead{Reflection\tablenotemark{d}}
}
\startdata

Si            & $1.74^{+0.20}_{-0.20}$ & 0.08  & 1.2  & 0.23  &18.6 & 0.18 \\ 
S             & $2.34^{+0.04}_{-0.07}$ & 0.26  & 17  & 0.30  &22.4 & 0.17 \\ 
Ar            & $3.05^{+0.06}_{-0.05}$ & 0.062 & 3.4  & 0.08  & 9.2 & 0.066 \\ 
Ca            & $3.73^{+0.08}_{-0.10}$ & 0.018 & 0.06 & 0.067 & 1.3 & 0.051 \\ 
Fe (K$\alpha$)& $6.49^{+0.01}_{-0.01}$ & 3.1   & 520  & 2.9   & 540 & 3.0 \\
Fe (K$\beta $)& $7.18^{+0.04}_{-0.05}$ & 0.42  &  59  & 0.40  &  49 & 0.38 \\
Ni            & $7.56^{+0.12}_{-0.14}$ & 0.038 & 1.3  & 0.013 & 3.5 & 0.10

\enddata
\tablenotetext{a}{Derived fluorescent line intensities when the highly
  ionized line emission component is modeled as plasmas in \cie\ and \pie.}
\tablenotetext{b}{Measured line centroids in the \pie\ fit, which are fixed
  in the \cie\ fit.}
\tablenotetext{c}{In multiples of $10^{-4} ~\rm{photons~cm}^{-2} ~\rm{s}^{-1}$}
\tablenotetext{d}{Line intensities expected from pure cold reflection
  normalized to the average observed Fe K$\alpha$ flux.}
\end{deluxetable}

\begin{deluxetable}{ll}
\tablecolumns{2}
\tablewidth{0pt}
\tablecaption{Summary of the Collisional Ionization Equilibrium Fit
 \label{tbl2}}
\tablehead{
  \colhead{Parameter} &
  \colhead{Value}
}

\startdata
$N_H$                    & $1.4 \times 10^{22} ~\rm{cm}^{-2}$ \\
$\Gamma$ \tablenotemark{a}        & 1.5 \\
$L_{\rm{soft}}$ \tablenotemark{b} & $3 \times 10^{40} ~\rm{erg~s}^{-1}$ \\
$kT$                     & $0.69^{+0.05}_{-0.07}$ keV \\
$EM$                     & $(1.12^{+0.17}_{-0.10}) \times 10^{63} ~\rm{cm}^{-3}$
\enddata
\tablenotetext{a}{Photon index of the scattered and reflected continuum
                  components, which are set equal in the fit.}
\tablenotetext{b}{Derived luminosity in the soft continuum component.}
\end{deluxetable}

\begin{deluxetable}{lcc}
\tablecolumns{2}
\tablewidth{0pt}
\tablecaption{Continuum Parameters for the Photoionization Equilibrium Fit
 \label{tbl3}}
\tablehead{
  \colhead{} & \multicolumn{2}{c}{Values} \\
  \cline{2-3} \\
  \colhead{Parameter} &
  \colhead{Empirical} &
  \colhead{DEM}
}

\startdata

$N_H$ ($10^{22}~\rm{cm}^{-2}$) & 0.87 & 0.89 \\
$\Gamma$ \tablenotemark{a}     & 1.8 & 1.6 \\
$L_{\rm{soft}}$ ($\rm{erg~s}^{-1}$) & $1.2 \times 10^{40}$ & $8.4 \times 10^{39}$

\enddata
\tablenotetext{a}{Photon index of the scattered and reflected continuum
                  components, which are set equal in the fit.}
\end{deluxetable}

\begin{deluxetable}{ll}
\tablecolumns{2}
\tablewidth{0pt}
\tablecaption{Summary of the Photoionization Ionization Equilibrium Fit \label{tbl4}}
\tablehead{
  \colhead{Parameter} &
  \colhead{Value\tablenotemark{a}}
}

\startdata
$\beta$           & $1.48^{+0.03}_{-0.04}$ \\
$\xi_{\rm{min}}$  & $0.91^{+0.48}_{-\infty}$ \\
$\xi_{\rm{max}}$  & $3.53^{+0.32}_{-0.20}$ \\
$EM$              & $5.5 \pm 0.7 \times 10^{63} ~\rm{cm}^{-3}$ \\
\enddata
\tablenotetext{a}{Errors correspond to 90\% confidence ranges for two
                  interesting parameters as described in the text.}
\end{deluxetable}

\end{document}